\def \farcs{\hbox{$.\!\!^{\prime\prime}$}}

\documentstyle[11pt,paspconf,epsf]{article}

\begin{document}

\title{Mass and mass-to-light ratio of galaxy groups from weak lensing}

\author{Henk Hoekstra$^1$, Marijn Franx$^2$, \& Konrad Kuijken$^1$}
\affil{	$^1$ Kapteyn Astronomical Institute\\
        P.O. Box 800, 9700 AV Groningen, The Netherlands\\
	e-mail: hoekstra, kuijken@astro.rug.nl\\
	$^2$ Leiden Observatory\\
 	P.O. Box 9513, 2300 RA Leiden, The Netherlands\\
	e-mail: franx@strw.strw.leidenuniv.nl}



\begin{abstract}
We have measured for the first time the weak lensing signal due to
groups of galaxies. The groups are at intermediate redshifts, and have
been identified in the CNOC2 field survey, which was kindly made available
by Ray Carlberg and Howard Yee. The ensemble averaged group
velocity dispersion, based on a preliminary selection of 59 groups, is
found to be $\langle\sigma^2\rangle^{1/2}= 320^{+46}_{-54}$ km/s,
which is in fair agreement with the dynamical estimate from the
spectroscopic redshifts.  Under the assumption that mass traces light,
we find an average mass-to-light ratio in restframe $B$ of
$(256\pm84)h {\rm M}_\odot/{\rm L}_{B\odot}$. Using this result we
obtain $\Omega_m=0.22\pm0.09~(\Omega_\Lambda=0)$.
\end{abstract}




\section{Introduction}

Galaxy groups, like the Local Group, are the most common structures in the 
universe. Despite being numerous, groups are hard to identify because the 
contrast with the smooth background of galaxies is low, and their galaxy 
properties are that of the field. To date most systems have been found using 
large redshift surveys or X-ray observations.

Measuring the mass locked up in these systems is important, but difficult 
(cf. Gott \& Turner 1977). Nolthenius \& White (1987) showed  that the 
masses inferred from redshift surveys depend on the survey parameters, the 
group selection procedure, and the way galaxies cluster. Consequently, an 
independent measure of the group mass is needed. Here we study the groups by 
their weak lensing effect on the shapes of the images of the faint background 
galaxies.

The weak lensing signal is maximal if the lenses are at intermediate 
redshifts, but even then, given the low masses of these systems,
the expected signals are too low to yield significant detections for
individual groups. Thus we have to study the ensemble averaged
signal of a large number of groups at intermediate redshifts.

The groups identified in the Canadian Network for Observational Cosmology 
Field Galaxy Redshift Survey (CNOC2) (e.g. Carlberg et al. 1998; Lin et al. 
1999) are ideal targets for our study. The aim of the CNOC2 survey is to study
the population of field galaxies at intermediate redshifts $(z=0.15-0.55)$. 
To do so, four widely separated patches on the sky were selected, for which 
multi-colour data were obtained, as well as spectroscopic redshifts for 
$\sim 5000$ galaxies brighter than $R_C=21.5$. The survey allows the 
identification of a large number of groups at intermediate redshifts.

\section{Data analysis}

We obtained deep $R$-band images of the central 31 by 23 arcminutes of two 
patches from the CNOC2 survey using the 4.2m William Herschel Telescope. To 
date, the combination of deep imaging an a large spectroscopic survey is 
unique, enabling us for the first time to study a large number of galaxy 
groups through weak lensing. A detailed discussion about the object analysis, 
including the corrections for the PSF can be found in Hoekstra et al. (1999b).
We end up with catalogues of $\sim 30000$ galaxies with $22<R<26$ in each
field. These galaxies are used to measure the weak lensing signal, enabling 
us to study the average properties of an ensemble of 59 groups from the CNOC2 
survey.

\section{Weak lensing analysis}

Most of the groups are relatively poor, and many of these
systems have been selected on the basis of only a few members. 
The first question that comes to mind is whether the selected structures
are genuine. The detection of a weak lensing signal provides an important
test to check the validity of the group selection.

Ideally one would like to scale the signals of the various groups
with an estimate of their mass, but the uncertainty in the observed velocity 
dispersions are too large. Therefore we assume that all groups have the
same mass and mass profile, and scale the signals of the various groups to 
that corresponding to the `average' group at $z=0.4$.

\begin{figure}
\begin{center}
\epsfxsize=\hsize
\epsfbox[0 175 600 690]{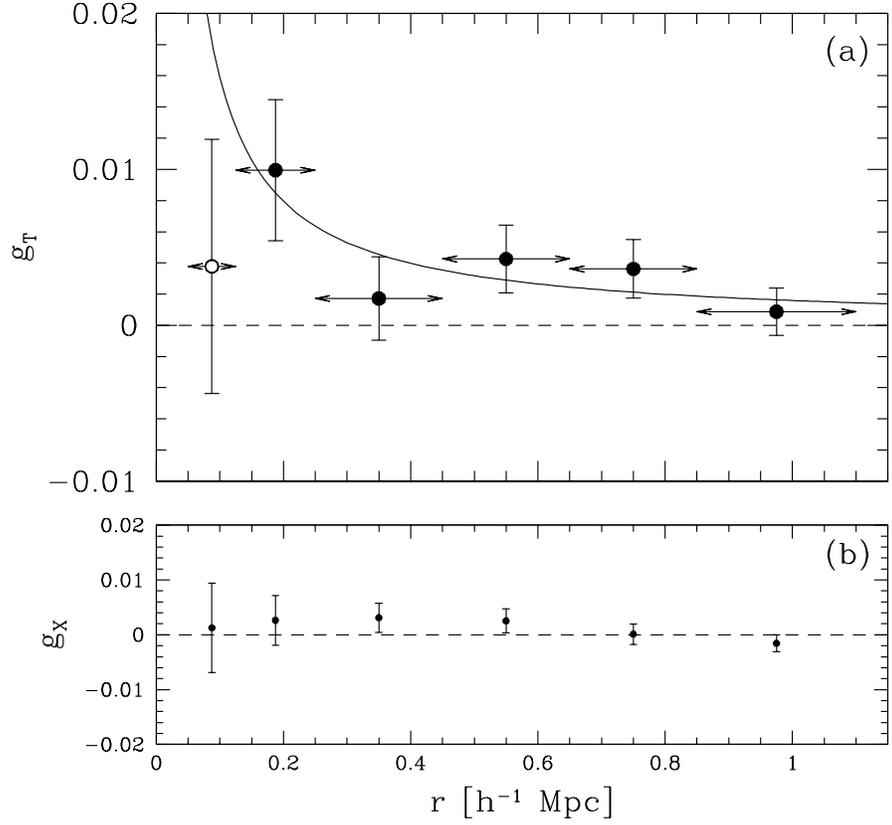}
\end{center}
\caption{Ensemble averaged tangential distortion as a function of radius 
around the 59 galaxy groups from the CNOC2 survey. The amplitude of the 
signal corresponds to that of the `average' group at a redshift of $z=0.4$. 
The solid line corresponds to the best fit singular isothermal sphere 
(SIS) model which has a velocity dispersion of 320~km/s. The 
lower panel shows the average signal when the phase of the distortion is 
increased by $\pi/2$, which should vanish if the observed signal is due to 
gravitational lensing. 
\label{fig:gtgroup}}
\end{figure}

Figure~\ref{fig:gtgroup} shows the ensemble averaged tangential distortion
as a function of radius around the 59 groups taken from the CNOC2 survey.
The amplitude of the signal, which is significant at the 99.8\% confidence
level, corresponds to that of the `average' group at a redshift of $z=0.4$. 
Various tests, like increasing the phase of the distortion by $\pi/2$,
placing the groups at random positions, or randomizing the ellipticities
of the sources yield no signal. Furthermore the results are robust
against imperfect corrections for the PSF anisotropy. We therefore conclude 
the detected signal is due to weak lensing by galaxy groups.

The lensing signal is detected out to a large distance from the group
centre. As the group members are believed to reside in a common group
halo, it is evident that the presence of the galaxy groups will
complicate attempts to constrain the sizes of halos of field galaxies.

Fitting a singular isothermal sphere model $(\kappa=r_E/2r)$ to the
observed distortion yields $r_E=0\farcs99\pm0\farcs30$. To relate this
measurement to an estimate of the average mass of the groups we use the 
photometric redshift distribution inferred from both Hubble Deep Fields 
(cf. Hoekstra et al. 1999a), converted to the $R$ band. As the groups are on 
average at relatively low redshifts, the dependence of the mass estimate 
on the redshift distribution is rather weak. 

Thus we find that the observed distortion corresponds to 
$\langle\sigma^2\rangle^{1/2}=320^{+46}_{-54}$ km/s. This result is in good
agreement with the dynamical estimate of $\langle\sigma^2\rangle^{1/2}=
251\pm21$ km/s, based on the group velocity dispersions.

\subsection{Mass-to-light ratio}

Under the assumption that the light traces the mass we computed the
lensing signal corresponding to the ensemble averaged light distribution 
and fitted this to the observed lensing signal. We do not observe a trend of 
the mass-to-light ratio with radius, and we find an average value of
$(256\pm84) h {\rm M}_\odot/{\rm L}_{B\odot}$ in the restframe $B$ band.
After correction for luminosity evolution (e.g. Lin et al. 1999) we find
a value of $(372\pm122) h {\rm M}_\odot/{\rm L}_{B\odot}$, which 
is somewhat lower than what is typically found for rich clusters of galaxies
(e.g. Carlberg et al. 1997).  

Similar to what is done for rich clusters of galaxies (e.g. Carlberg et al. 
1997; Carlberg et al. 1999) or galaxy groups (e.g. Gott \& Turner 1977), 
we can use our measurement of the group mass-to-light ratio to obtain an 
estimate of the density of the universe, for which we find
$\Omega_m=0.22\pm0.09$ taking $\Omega_\Lambda=0$ $(\Omega_m=0.14\pm0.06~
{\rm for~}\Omega_\Lambda=1)$. A detailed analysis will be given in Hoekstra 
et al. (1999b).

\section{Conclusions}

The detection of the weak lensing signal of a preliminary selection of
galaxy groups first of all shows that the CNOC2 survey allows the
identification of these systems at intermediate redshifts.  This is
supported even more by the agreement between the weak lensing, and
dynamical mass estimates. The ensemble averaged group velocity
dispersion, based on the 59 selected groups, is found to be
$\langle\sigma^2\rangle^{1/2}= 320^{+46}_{-54}$ km/s, which is in fair
agreement with the dynamical estimates.

Under the assumption that mass traces light, we find an average mass-to-light 
ratio in the restframe $B$ band of $(256\pm84) h 
{\rm M}_\odot/{\rm L}_{B\odot}$. This yields an estimate for $\Omega_m$ of
$0.22\pm0.09~(\Omega_\Lambda=0)$. A detailed discussion of the preliminary 
results presented here can be found in Hoekstra et al. (1999b).

\bigskip
\noindent {\it This work was done in collaboration with Ray Carlberg and 
Howard Yee.}

\end{document}